\documentstyle[prd,aps,multicol,epsfig]{revtex}
\newcommand\bq{\begin{equation}}
\newcommand\eq{\end{equation}}
\newcommand{\bqn}{\begin{eqnarray}}
\newcommand{\eqn}{\end{eqnarray}}

\begin{document}
\draft
\title{Stochastic background of gravitational waves}
\author{Jos\'e C.N. de Araujo\thanks{Email address: jcarlos@das.inpe.br},
Oswaldo D. Miranda\thanks{Email address: oswaldo@das.inpe.br} and
Odylio D. Aguiar\thanks{Email address: odylio@das.inpe.br}}
\address{Divis\~ao de Astrof\'\i sica, Instituto Nacional
de Pesquisas Espaciais,\\ Avenida dos Astronautas, 1758, S\~ao Jos\'e
dos Campos, S.P. 12227-010 , Brazil}
\date{\today}
\maketitle
\begin{abstract}

A continuous stochastic background of gravitational waves (GWs)
for burst sources is produced if the mean time interval between
the occurrence of bursts is smaller than the average time duration
of a single burst at the emission, i.e., the so called duty cycle
must be greater than one. To evaluate the background of GWs
produced by an ensemble of sources, during their formation, for
example, one needs to know the average energy flux emitted during
the formation of a single object and the formation rate of such
objects as well. In many cases the energy flux emitted during an
event of production of GWs is not known in detail, only
characteristic values for the dimensionless amplitude and
frequencies are known. Here we present a shortcut to calculate
stochastic backgrounds of GWs produced from cosmological sources.
For this approach it is not necessary to know in detail the energy
flux emitted at each frequency. Knowing the characteristic values
for the ``lumped'' dimensionless amplitude and frequency we show
that it is possible to calculate the stochastic background of GWs
produced by an ensemble of sources.
\end{abstract}


\begin{multicols}{2}

\section{Introduction}

The detection of gravitational radiation will probably mark a new
revolution in the history of astronomy. It is worth mentioning
that the detection of gravitational waves (GWs) will directly
verify the predictions of the general relativity theory concerning
the existence or not of such waves, as well as other theories of
gravity \cite{thor87}.

The realm of astrophysics is the place where one finds sources of
GWs detectable by the GW observatories. There is a host of
possible astrophysical sources of GWs: namely, supernovas, the
collapse of a star or star cluster to form a black hole, inspiral
and coalescence of compact binaries, the fall of stars and black
holes into supermassive black holes, rotating neutron stars,
ordinary binary stars, relics of the big bang, vibrating or
colliding of monopoles, cosmic strings, etc., among others
\cite{thor87}. Nowadays there is a great effort to study, from the
theoretical point of view, which are the most promising sources of
GWs to be detected, in particular, their wave forms,
characteristic frequencies, and the number of sources a year that
one expects to observe.

In a few years, instead of building models trying to understand
how the sources of GWs work, it will be possible, starting from
the observations (wave forms, amplitudes, polarizations, etc.) to
really understand how the GW emission takes place.

Because of the fact the GWs are produced by a large variety of
astrophysical sources and cosmological phenomena it is quite
probable that the Universe is pervaded by a background of such
waves. Binary stars of a variety of stars (ordinary, compact, or
combinations of them), population III stars, phase transitions in
the early Universe, and cosmic strings are examples of sources
that could generate such putative background of GWs.

As the GWs possess a very weak interaction with matter passing
through it with impunity, relic radiation (spectral properties,
for example) once detected can provide information on the physical
conditions from the era in which the GWs were produced. In
principle it will be possible, for example, to get information
from the epoch when the galaxies and stars started to form and
evolve.

Here we present, in particular, a shortcut to the calculation of
stochastic background of GWs. For this approach it is not
necessary to know in detail the energy flux of the GWs produced in
a given burst event. If the characteristic values for the
dimensionless amplitude and frequency are known and the event rate
is given it is possible to calculate the stochastic background of
GWs produced by an ensemble of sources of the same kind.

This paper is organized as follows. In Sec. II we show how to
calculate the stochastic background of GWs starting from
characteristic values for the dimensionless amplitude and
frequency as well as the burst event rate. In Sec. III we apply
the idea presented in Sec. II to the calculation of a stochastic
background of GWs from a cosmological population of black holes,
and finally in Sec. IV we present the conclusions.

\section{A Shortcut to the calculation of stochastic background
of GW{\bf\scriptsize s}}

        The GWs can be characterized by their
dimensionless amplitude $h$, and frequency $\nu$. The spectral
energy density, the flux of GWs, received on Earth, $F_\nu$, in
${\rm erg}\; {\rm cm}^{-2}{\rm s}^{-1}{\rm Hz}^{-1}$, is (see,
e.g., Refs. \cite{doug79,hils90})

\begin{equation}
F_{\nu} = {c^{3}s_{\rm h}\omega_{\rm obs}^{2}\over 16\pi G},
\end{equation}

\noindent where $\omega_{\rm obs}=2\pi \nu_{\rm obs}$ with
$\nu_{\rm obs}$ the GW frequency observed on Earth (in ${\rm
Hz}$), $c$ is the speed of light, $G$ is the gravitational
constant, and $\sqrt{s_{\rm h}}$ is the strain amplitude of the GW
(in $\rm Hz^{-1/2}$). For $\omega \geq 0$, Eq.(1) must be
multiplied by a factor of 2 in order to account for the folding of
negative frequencies into positive (see, e.g., Ref.
\cite{ferr99a}).
        The stochastic background produced by an ensemble of sources,
of the same kind, would have a spectral density of the flux and strain
amplitude also related to the above equation. The strain amplitude
at a given frequency at the present time could be, for example, a
contribution of sources of the same kind but with different masses
producing GWs at different redshifts. Thus, the ensemble of sources
produces a background whose characteristic amplitude at the present
time is $\sqrt s_{\rm h}$.

    On the other hand, the spectral density of the flux can be written
as (see, e.g., Ref. \cite{ferr99a,ferr99b})

\begin{equation}
F_{\nu}=\int f_{\nu}(\nu_{\rm obs})dR,
\end{equation}

\noindent where $f_{\nu}(\nu_{\rm obs})$ is the energy flux per
unit of frequency (in ${\rm erg}\;{\rm cm}^{-2}\;{\rm Hz}^{-1}$)
produced by a unique source and $dR$ is the differential rate of
production of GWs by the source.

        The energy flux per unit frequency $f_{\nu}(\nu_{\rm obs})$
can be written as follows (see, e.g., Ref. \cite{carr80})

\begin{equation}
f_{\nu}(\nu_{\rm obs}) = {\pi c^{3}\over 2G}h_{\rm single}^{2},
\end{equation}

\noindent where $h_{\rm single}$ is the dimensionless amplitude produced by
an event that generates a signal with observed frequency $\nu_{\rm obs}$.

    Then, the resulting equation for the spectral density of the flux is

\begin{equation}
F_{\nu} = {\pi c^{3}\over 2G} \int h_{\rm single}^{2}dR.
\end{equation}

        From the above equations we obtain for the strain

\begin{equation}
s_{\rm h} = {1 \over \nu_{\rm obs}^{2}}\int h_{\rm single}^{2} dR.
\end{equation}

        Thus, the dimensionless amplitude reads

\begin{equation}
h_{\rm BG}^{2} = {1 \over \nu_{\rm obs}}\int h_{\rm single}^{2} dR.
\end{equation}

        With the above equations one finds, for example, the dimensionless
amplitude of the GWs produced by an ensemble of sources of the
same kind that generates a signal observed at frequency $\nu_{\rm
obs}$. Note that in this formulation it is not necessary to know
in detail the energy flux of GWs at each frequency. Knowing the
characteristic amplitude for a given source, $h_{\rm single}$,
associated to an event burst of GWs, and the rate of production of
GWs, it is possible to obtain the stochastic background of an
ensemble of these sources.

It is worth mentioning that if the collective effect of bursts of
GWs really form a continuous background the quantity called duty
cycle must be greater than one. In other words: the mean time
interval between the occurrence of bursts must be smaller than the
typical duration of each burst. The duty cycle is defined as
follows:

\begin{equation}
D(z)=\int dR\bar{\Delta\tau_{\rm GW}}(1+z),
\end{equation}

\noindent where $\bar{\Delta\tau_{\rm GW}}$ is the average time
duration of single bursts at the emission (see, e.g., Ref.
\cite{ferr99a}).

In the present study we are using the relationship between
$h_{BG}$  and $h_{single}$ (Eq. 6) as Ferrari {\it et al.}
\cite{ferr99a,ferr99b}, in either case of duty cycle: large or
small. In the next section we apply the technique for a case in
which the duty cycle is small; in another study, to appear
elsewhere, we apply it to a case where the duty cycle is large
(see Ref. \cite{mira99a}).

        The energy density of GWs is usually written in terms of the closure
energy density of GWs per logarithmic frequency interval, which is
given by

\begin{equation}
\Omega_{\rm GW} = {1\over \rho_{\rm c}} {d\rho_{\rm GW}\over d\log \nu_{\rm obs}},
\end{equation}

\noindent where $\rho_{\rm c}$ is the critical density ($\rho_{\rm c}=3H^{2}/8\pi G$).
The above can be written as

\begin{equation}
\Omega_{\rm GW} = {\nu_{\rm obs}\over c^{3}\rho_{\rm c}}F_{\nu} =
{4\pi^{2}\over 3H^{2}}\nu_{\rm obs}^{2} h_{\rm obs}^{2}.
\end{equation}

\section{Application: stochastic background of GW{\bf\scriptsize s}
from a cosmological population of stellar black holes}

In this section we apply the formulation presented in the precedent section
to calculate the background of GWs from a cosmological population of
stellar black holes.

From Eq. (6) one sees that it is necessary to know (a) $h_{\rm
single}$, here named $h_{\rm BH}$, the characteristic amplitude of
the burst of GWs produced during the black hole formation; (b)
$dR$, the differential rate of production of GWs, here named
$dR_{\rm BH}$, the differential rate of black hole formation. It
is worth noting that we are implicitly assuming that during the
formation of each black hole there is a production of a burst of
GWs.

To proceed it is necessary to know the star formation history of
the Universe, which we adopted from a study performed by Madau
{\it et al.} \cite{mada98a}, which holds for the redshift range $0
< z < 5$. It is also necessary to know (a) the initial mass
function (IMF), which we assume to be the Salpeter IMF, and (b)
the smallest progenitor mass which is expected to lead to black
holes (see Refs. \cite{timm95,woos96}).

\subsection{The rate of stellar black holes formation}

        The differential rate of black hole formation can be written as

\begin{equation}
dR_{\rm BH} = \dot\rho_{\ast}(z) {dV\over dz} \phi(m)dmdz,
\end{equation}

\noindent where $\dot\rho_{\ast}(z)$ is the star formation rate
(SFR) density (in $\rm M_{\odot}yr^{-1}Mpc^{-3}$), $dV$ is the
comoving volume element, and $\phi(m)$ the IMF (see
Refs.\cite{ferr99a,ferr99b,mira99a}).

The SFR density can be derived from observations.
In particular, our present view of the Universe at
redshifts $z\;{^<_\sim}\;4-5$ has been extended by recent data
obtained with the Hubble Space Telescope (HST) and other large
telescopes (see e.g. Refs. \cite{lill96,mada96,elli97}).

    It has been shown that, in general, the measured comoving luminosity
density is proportional to the SFR density. Thus, the star
formation evolution can be derived from recent UV-optical
observations of star forming galaxies out to redshifts $\sim 4-5$
\cite{mada98b}. Figure 1 shows  the SFR density obtained by  Madau
{\it et al.} \cite{mada98b}.

\begin{figure}
\begin{center}
\leavevmode
\centerline{\epsfig{figure=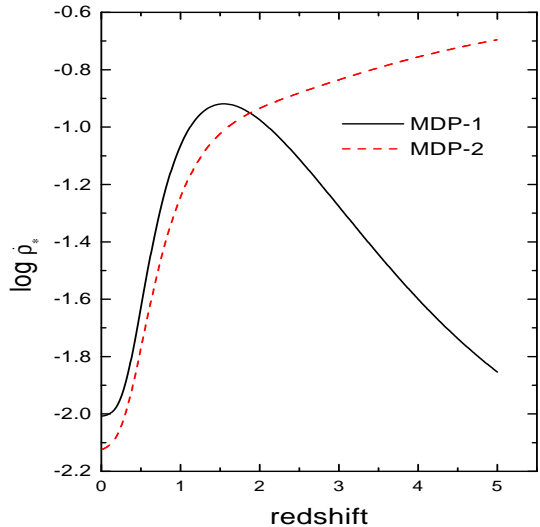,angle=360,height=8.5cm,width=8.5cm}}
\caption{Evolution of the log of the SFR density
($\rm M_{\odot}{yr}^{-1}{Mpc}^{-3}$) for $\Omega_0=1$,
($\Lambda=0$), $h = 0.5$ and a Salpeter IMF. The solid
line represents the SFR density evolution given by Eq. (11), MDP-1,
beyond the dotted line corresponds to the SFR density given by
Eq. (12), MDP-2.}
\end{center}
\label{fig1}
\end{figure}
\begin{figure}
\begin{center}
\leavevmode
\centerline{\epsfig{figure=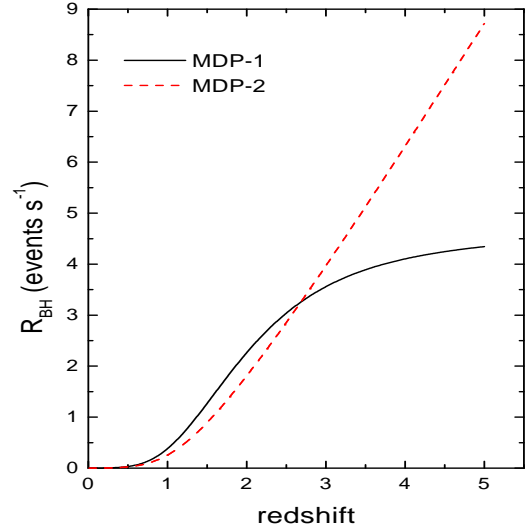,angle=360,height=8.5cm,width=8.5cm}}
\caption{Evolution of the rate of black hole formation occurring per unit time within
the comoving volume out to redshift $z$ for $\Omega_0=1$, ($\Lambda=0$), $h = 0.5$ and
a Salpeter IMF. The solid line represents the rate of black hole formation when we used
the Eq. (11) to the SFR density, MDP-1, beyond the dotted line corresponds to the rate
of black hole formation when we used the SFR density given by Eq. (12), MDP-2, (see
Fig. 1).}
\end{center}
\label{fig2}
\end{figure}

        In particular, there are two different fits to the SFR density
presented by these authors. The first fit for the SFR density
(here after referred to as MDP-1) is given by

\begin{eqnarray}
\dot\rho_{\ast}(z)= 0.049\,[t_{9}^{5}\,{\rm e}^{-t_{9}/0.64} +
\; 0.2(1 - {\rm e}^{-t_{9}/0.64})] \nonumber \\
\times { \rm M_{\odot}\; yr^{-1}Mpc^{-3}},\qquad\qquad\qquad\qquad
\end{eqnarray}

\noindent where $t_{9}$ is the Hubble time in $\rm Gyr$ [$t_{9}=13/(1+z)^{3/2}$].

        The second fit for the SFR density (here after referred to as MDP-2)
is given by

\begin{eqnarray}
\dot\rho_{\ast}(z) = 0.336\,{\rm e}^{-t_{9}/1.6} + \; 0.0074(1-{\rm e}^{-t_{9}/0.64})
\nonumber \\
 + 0.0197\,t_{9}^{5}\,{\rm e}^{-t_{9}/0.64} \;\; {\rm M_{\odot}\; yr^{-1}Mpc^{-3}}.
\end{eqnarray}

        In the above fits, Eqs. (11) and (12), Madau and collaborators
considered an Einstein - de Sitter cosmology ($\Omega_{0}=1$) with Hubble
constant $H_{0} = 50\; {\rm km\; s^{-1}Mpc^{-1}}$ and cosmological constant
$\Lambda = 0$. Note that for a different cosmological scenario it is necessary
to rescale the SFR density.

    The fit given by Eq. (11) traces the rise, peak, and sharp drop
of the observed UV emissivity at redshifts $z\;{^>_\sim}\; 2$,
while the fit given by Eq. (12) considers that half of the
present-day stars, the fraction contained in spheroidal systems
\cite{sche87}, were formed at $z>2.5$ and were enshrouded by dust.
This fact produces an increase in the SFR density at redshifts
$z>2.5$ (see Fig. 1) contrary to the sharp drop described in
Eq. (11).

    The consistency of $\dot\rho_{\ast}(z)$ given by Eq. (12) with the Hubble
Deep Field (HDF) analysis is obtained assuming a dust extinction
that increases with redshift. This fact is consistent with the
evolution of the luminosity density, but overpredicts the metal
mass density at high redshifts as derived from quasar absorbers (see
Ref. \cite{mada98a}).

    Despite this fact, it is interesting also to analyze the GW
production with $\dot\rho_{\ast}(z)$ given by MDP-2 [Eq. (12)]
because this SFR density produces a large number of supernovas at
$z>2.5$, when compared to the SFR history described by MDP-1.

        Concerning the IMF here we consider Salpeter's, as already
mentioned. Thus,

\begin{equation}
\phi(m) = A m^{-(1+x)},
\end{equation}

\noindent where $A$ is a normalization constant and $x=1.35$ the
Salpeter exponent.

    The IMF is defined in such a way that $\phi(m)dm$ represents
the number of stars in the mass interval $[m,m+dm]$. The
normalization of the IMF is obtained through the relation

\begin{equation}
\int_{\rm m_l}^{\rm m_u} m\phi(m)dm = 1,
\end{equation}

\noindent with ${\rm m_{l}} = 0.1 {\rm M}_{\odot}$ and ${\rm
m_{u}}=125 {\rm M}_{\odot}$. Using this normalization of the mass
spectrum, we obtain $A=0.17\; ({\rm M}_{\odot})^{0.35}$.

    In the present work we follow Timmes, Woosley \& Wheaver \cite{timm95}
(see also Ref. \cite{woos96}), who  obtain from stellar evolution
calculations, the minimal progenitor mass to form black holes,
namely, $18{\rm M}_{\odot}$ to $30{\rm M}_{\odot}$ depending on
the iron core masses. Then, we assume that the minimum mass able
to form a remnant black hole is $m_{\rm min}=25 {\rm M}_{\odot}$.
For the remnant mass, $M_{\rm r}$, we take $M_{\rm r}=\alpha m$,
where $m$ is the mass of the progenitor star and $\alpha = 0.1$
(see, e.g., Refs. \cite{ferr99a,ferr99b}).

In Fig. 2 we show the evolution of the rate of black hole
formation $R_{\rm BH}(z)$, i.e., the number of black holes formed
per unit time within the comoving volume out to redshift $z$, for
MDP-1 and MDP-2 for a cosmological scenario with $\Omega_{0} =1.0$
and $h_{0}=0.5$. Note that MDP1 and MDP-2 are similar for $z <
2.5$, and for $z > 2.5$ they are quite different.

\subsection{The gravitational wave production}

    To obtain the stochastic background, besides knowing the differential
rate of black holes formation presented in Sec. II, one needs to
know $h_{\rm BH}$, the characteristic dimensionless amplitude
generated during the black hole formation. Following Thorne
\cite{thor87}, $h_{\rm BH}$ reads

\begin{eqnarray}
h_{\rm BH}\lefteqn{=\bigg({15\over 2\pi}\varepsilon\bigg)^{1/2}{G\over
c^{2}} {M_{\rm r}\over r_{\rm z}} {}} \nonumber \\ & & {}\simeq 7.4\times
10^{-20}\varepsilon^{1/2}\bigg({M_{\rm r}\over {\rm
M}_{\odot}}\bigg) \bigg({r_{\rm z}\over 1{\rm Mpc}}\bigg)^{-1},
\end{eqnarray}

\noindent where $\varepsilon$ is the efficiency of generation of
GWs and $r_{\rm z}$ is the distance to the source.

The collapse to a black hole produces a signal with frequency
(see, e.g., Ref. \cite{thor87})

\begin{eqnarray}
\nu_{\rm{obs}} \lefteqn{=  {1\over 5{\rm \pi} M_{\rm r}}{c^{3}\over
G}(1+z)^{-1} {} } \nonumber\\ & & {} \simeq 1.3\times 10^{4}{\rm
Hz}\bigg({{\rm M}_{\odot}\over M_{\rm r}}\bigg)(1+z)^{-1},
\end{eqnarray}

\noindent where the factor $(1+z)^{-1}$ takes into account the
redshift effect on the emission frequency, that is, a signal
emitted at frequency $\nu_{\rm e}$ at redshift $z$ is observed at
frequency $\nu_{\rm obs}=\nu_{\rm e}(1+z)^{-1}$.

        From Eqs. (6), (10) and (15) we obtain, for the dimensionless
amplitude (for $\alpha = 0.1$),

\begin{eqnarray}
h_{\rm BG}^{2}={(7.4\times 10^{-21})^{2}\varepsilon \over \nu_{\rm
obs}} \bigg[\int_{z_{\rm c_f}}^{z_{\rm c_i}} \int_{m_{\rm
min}}^{m_{\rm u}} \bigg({m\over {\rm M}_{\odot}}
\bigg)^{2}\bigg({d_{L}\over 1{\rm Mpc}} \bigg)^{-2} \nonumber
\end{eqnarray}
\begin{equation}
\times \;\dot\rho_{\ast}(z){dV\over dz}
\phi(m)dmdz\bigg],\qquad\qquad\qquad
\end{equation}

\noindent where in the above equation $d_{\rm L}$ is the luminosity distance
to the source.

    The comoving volume element is given by

\begin{equation}
dV = 4\pi\bigg({c\over H_{0}}\bigg) r_{\rm z}^{2} {dz\over (1+z)},
\end{equation}

\noindent and the comoving distance, $r_{\rm z}$, is

\begin{equation}
r_{\rm z}={2c[1-(1+z)^{-1/2}]\over H_{0}}.
\end{equation}

    In the above equation the density parameter is considered $\Omega_{0}=1$ and
$H_{0}$ is the present value of the Hubble parameter.

    The comoving distance is related to the luminosity distance by

\begin{equation}
d_{\rm L} = r_{\rm z} (1+z)
\end{equation}

With the above equations we can calculate the dimensionless
amplitude produced by an ensemble of black holes that generates a
signal observed at frequency $\nu_{\rm obs}$.

It is worth mentioning  that the formulation used here is similar
to that used by Ferrari {\it et al.} \cite{ferr99a,ferr99b}, but
instead of using an average energy flux taken from Stark \& Piran
\cite{star86}, who simulated the axisymmetric collapse of a
rotating polytropic star to a black hole, we use Eq. (15) to
obtain the energy flux, which takes into account the most relevant
quasinormal modes of a rotating black hole and represents a kind
of average over the rotational parameter. Both formulations
present similar results, since in the end the most important
contributions to the energy flux come from the quasinormal modes
of the black hole formed, which account for most of the
gravitational radiation produced during the collapse process. In a
paper to appear elsewhere \cite{mira99b} we present a detailed
comparison between our formulation and results with those by
Ferrari {\it et al.} \cite{ferr99a}.

\subsection{Numerical results}

        Fig. 3 presents the amplitude of GWs as a function of the
observed frequency obtained from Eq. (17) for the two SFR
densities present in Sec. II. We obtained that the star formation
rate given by Eq. (12), MDP-2, produces a maximum amplitude
$h_{\rm BG}$ lower than the MDP-1 SFR density. This seems to be a
contradiction since $R_{\rm BH}$ is higher for MDP-2. Note,
however, that for $z < 2.5$, $R_{\rm BH}$ is higher for MDP-1 and
due to this fact the maximum amplitude peak is higher for  MDP-1.
The SFR density described by MDP-2 produces a higher $R_{\rm BH}$
for $z > 2.5$, but the contribution of these events do not
contribute to enhance the $h_{\rm BG}$ peak, but instead
contribute to enhance $h_{\rm BG}$ at the lowest frequencies due
to the redshift effect (see also Fig. 2).

    A comparison of our results with those of Ferrari {\it et al.} \cite{ferr99a}
shows that the formulation used here present similar results. It
is worth mentioning that in the comparison we have adopted
$\varepsilon\sim 10^{-4}$. Note that in the present study, instead
of using the average energy flux emitted during the axisymmeric
collapse of a rotating polytropic star to a black hole with
different values for the rotational parameter, we use Eq. (15). In
this equation there is no explicit dependence of $h_{\rm BH}$ on
the rotational parameter, it represents a kind of characteristic
value for the amplitude of GWs during the black hole formation.
The characteristic frequency given by Eq. (16) has to do with the
frequency of the lowest $m=0$ quasinormal mode of a black hole,
which is believed to be excited during its formation.

    One could argue that it is surprising that our results agree so
well with those of Ferrari {\it et al.}, particularly those of
Fig. 3. The reason for this good agreement is related to the fact
that the main contribution to the strain amplitude in the Ferrari
{\it et al.} calculations comes from the lowest quasinormal mode
of the black hole formed, which is also the main contribution
present in the Eq. (15) we use in our study. A close comparison
shows, however, that the peak of the curve in the Fig. 3 occurs
for a frequency higher than that of Ferrari {\it et al.} Although
most of the energy comes from the quasinormal mode, there is a
contribution from the lower frequencies of the energy flux (see
Ref. \cite{ferr99a}), which moves the peak of the strain amplitude
of Ferrari {\it et al.} to the left as compared to ours.

\begin{figure}
\begin{center}
\leavevmode
\centerline{\epsfig{figure=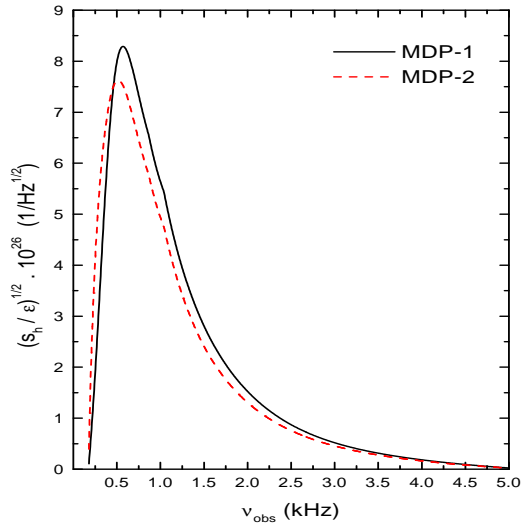,angle=360,height=8.5cm,width=8.5cm}}
\caption{The background amplitude of the GWs as a function of
$\nu_{\rm{obs}}$. The solid line (dashed line) represents
$(s_{\rm h}/\varepsilon)^{1/2}$ for the MDP-1 SFR density
(MDP-2 SFR density).}
\end{center}
\label{fig3}
\end{figure}

    We show in Fig. 4  $\Omega_{\rm GW}$ as a function of the observed
frequency. Note that the MDP-1 SFR density presents, due to the
higher contribution to $R_{\rm BH}$ for $z < 2.5$, a  more
relevant contribution to the closure energy density of GWs for
almost all frequencies, than the MDP-2 SFR density does.

    Comparing Fig. 4 with the corresponding figures of Ferrari {\it et al.},
one notes that their curves are broader than ours. This occurs due
to the fact the closure energy density is  directly proportional
to the energy flux, and therefore more sensitive to their
frequency dependence. The Ferrari {\it et al.} energy flux as a
function of frequency is broader than we use here, this is why
their closure energy density as function of frequency curves are
broader than ours.

    Certainly, modifying the exponent $x$ in Eq. (13) to $x>1.35$, we
will obtain a more steeply falling IMF, corresponding to a lower
number of massive stars than that obtained using Salpeter's IMF.
This produces, as a result, a lower rate of black holes formation
than that obtained here, narrowing the curves present in Figs. 3
and 4. However, the agreement with the study performed by Ferrari
{\it et al.} will be still good , since that their study present
the same dependence to the IMF. Thus both results (and models)
will be modify in the same way.

\begin{figure}
\begin{center}
\leavevmode
\centerline{\epsfig{figure=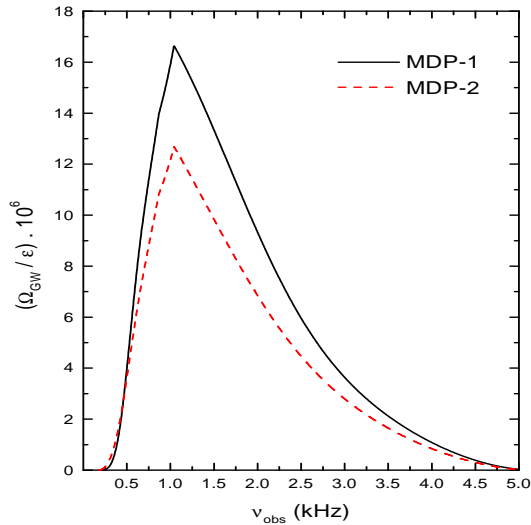,angle=360,height=8.5cm,width=8.5cm}}
\caption{The closure energy density of GWs as a function of
$\nu_{\rm{obs}}$. The solid line (dashed line) represents
$\Omega_{\rm GW}/\varepsilon$ for the MDP-1 SFR density (MDP-2 SFR density).}
\end{center}
\label{fig4}
\end{figure}

\section{Conclusions}

Here we present a shortcut to the calculation of stochastic
background of GWs. For this approach it is not necessary to know
in detail the energy flux at each frequency of the GWs produced in
a given burst event, if the characteristic values for the
``lumped'' dimensionless amplitude and frequency are known, and
the event rate is given, it is possible to calculate the
stochastic background of  GWs produced by an ensemble of sources
of the same kind.

   Since one knows the dominant processes of GWs emission one can calculate
the stochastic background of an ensemble of black holes. We argue
that the same holds for other processes of GWs production,
particularly those involving cosmological sources, since the
number of sources could be big enough to produce stochastic
backgrounds.

    We apply this formulation to the study of a stochastic background
of GWs produced during the formation of a cosmological population
of stellar black holes. We compare the results obtained here with
a study by Ferrari {\it et al.} \cite{ferr99a}, who take into
account in their calculations an average energy flux for the GWs
emitted during the formation of black holes obtained from
simulations by Stark \& Piran \cite{star86}. Our results are in
good agreement.

For most sources of GWs only characteristics values for the
dimensionless amplitude and frequency are known, if these sources
are numerous, a stochastic background of GWs could be produced. We
argue that the formulation  presented here could be applied to
other calculations of stochastic backgrounds as well.

\acknowledgments

JCNA would like to thank the Brazilian agency FAPESP for support
(grants 97/06024-4 and 97/13720-7). ODM would like to thank the
Brazilian agency FAPESP for support (grant 98/13735-7), and Dra.
Sueli Viegas for her continuous encouragement to the development
of this work. ODA thanks CNPq (Brazil) for financial support
(grant 300619/92-8). Finally, we would like to thank an anonymous
referee who has given useful suggestions that improved the present
version of our paper.

\label{lastpage}
\end{multicols}

\begin{thebibliography}{99}

\bibitem{thor87} K.S. Thorne , in {\it 300 years of Gravitation},
edited by S.W. Hawking and W. Israel (Cambridge University Press,
Cambridge, England, 1987), p.331.

\bibitem{doug79} D.H. Douglass and V.G. Braginsky, in {\it General Relativity:
An Einstein Centenary Survey}, edited by  S.W. Hawking and W.
Israel (Cambridge University Press, Cambridge, England, 1979),
p.90.

\bibitem{hils90} D. Hils, P.L. Bender, and R.F. Webbink, Astrophys. J. {\bf 360},
75 (1990).

\bibitem{ferr99a} V. Ferrari, S. Matarrese, and R. Schneider, Mon. Not. R. Astron. Soc.
{\bf 303}, 247 (1999).

\bibitem{ferr99b} V. Ferrari, S. Matarrese, and R. Schneider, Mon. Not. R. Astron. Soc.
{\bf 303}, 258 (1999).

\bibitem{carr80} B.J. Carr, Astron. Astrophys. {\bf 89}, 6 (1980).

\bibitem{mira99a} O.D. Miranda, J.C.N. de Araujo, and O.D. Aguiar
(unpublished).

\bibitem{mada98a} P. Madau, M. Della Valle, and N. Panagia,
Mon. Not. R. Astron. Soc.  {\bf 297}, L17 (1998).

\bibitem{timm95} F.X. Timmes, S.E. Woosley, and T.A. Weaver, Astrophys. J.,
Suppl. Ser. {\bf 98}, 617 (1995)

\bibitem{woos96} S.E. Woosley and F.X. Timmes, Nucl. Phys. {\bf A606},
137 (1996)

\bibitem{lill96} S.J. Lilly, O. Le F\'evre, F. Hammer, and D. Crampton,
Astrophys. J. Lett.  {\bf 460}, L1 (1996).

\bibitem{mada96} P. Madau, H.C. Ferguson, M.E. Dickinson, M. Giavalisco, C.C.
Steidel, and A. Fruchter, Mon. Not. R. Astron. Soc. {\bf 283},
1388 (1966).

\bibitem{elli97} R.S. Ellis, Annu. Rev. Astron. Astrophys. {\bf 35}, 389
(1997).

\bibitem{mada98b} P. Madau, L. Pozzetti, and M.E. Dickison, Astrophys. J. {\bf
498}, 106 (1998).

\bibitem{sche87}  P.L. Schechter and A. Dressler, Astron. J. {\bf 94}, 56
(1987).

\bibitem{star86} R.F. Stark and T. Piran, 1986, in {\it Proceedings of the Fourth
Marcel Grossmann Meeting on General Relativity}, edited by R.
Ruffini (Elsevier Science, Amsterdam, 1986), p.327.

\bibitem{mira99b} O.D. Miranda, J.C.N. de Araujo, and O.D. Aguiar
(in preparation).

\end{thebibliography}
\end{document}